\documentclass[pre,aps,floats,twocolumn,superscriptaddress,floatfix]{revtex4}
\usepackage{graphicx,amssymb,amsmath,ifthen}
\usepackage[active]{srcltx}

\begin{document}
\title{Taxonomy of particles in Ising spin chains}  
\author{Dan Liu}
\affiliation{
  Department of Physics,
  University of Rhode Island,
  Kingston RI 02881, USA}
  \author{Ping Lu}
\affiliation{
  Department of Physics,
  University of Rhode Island,
  Kingston RI 02881, USA}
\author{Gerhard M{\"{u}}ller}
\affiliation{
  Department of Physics,
  University of Rhode Island,
  Kingston RI 02881, USA}
\author{Michael Karbach}
\affiliation{
  Bergische Universit{\"{a}}t Wuppertal,
  Fachgruppe Physik,
  D-42097 Wuppertal, Germany}
\begin{abstract}
  The statistical mechanics of particles with shapes on a one-dimensional lattice is investigated 
  in the context of the $s=1$ Ising chain with uniform nearest-neighbor coupling, quadratic 
  single-site potential, and magnetic field, which supports four distinct ground states: 
  $|\uparrow\downarrow\uparrow\downarrow\cdots\rangle$, $|\circ\circ\cdots\rangle$, 
  $|\uparrow\uparrow\cdots\rangle$, $|\uparrow\circ\uparrow\circ\cdots\rangle$. 
  The complete spectrum is generated from each ground state by particles from a different set of
  six or seven species.
  Particles and elements of pseudo-vacuum are characterized by motifs (patterns of several 
  consecutive site variables). 
  Particles are floating objects that can be placed into open slots on the lattice. 
  Open slots are recognized as permissible links between motifs.
  The energy of a particle varies between species but is independent of where it is placed.
  Placement of one particle changes the open-slot configuration for particles of all species.
  This statistical interaction is encoded in a generalized Pauli principle, from which 
  the multiplicity of states for a given particle combination is determined and used for the 
  exact statistical mechanical analysis. 
  Particles from all species belong to one of four categories: compacts, hosts, tags, or hybrids. 
  Compacts and hosts find open slots in segments of pseudo-vacuum. 
  Tags find open slots inside hosts. 
  Hybrids are tags with hosting capability.  
  In the taxonomy of particles proposed here, `species' is indicative of structure and `category'
  indicative of function.
  The hosting function splits the Pauli principle into exclusion and accommodation parts.
  Near phase boundaries, the state of the Ising chain at low temperature is akin to that of 
  miscible or immiscible liquids with particles from one species acting as surfactant molecules.
\end{abstract}

\maketitle

%
\section{Introduction}\label{sec:intro}
%
The constituent particles of condensed matter are subject to strong 
interaction forces. 
Inspired by the search for separable coordinates in Hamiltonian mechanics,
much progress in the understanding of material properties
has resulted from finding collective modes that are much more weakly interacting.
The \mbox{(quasi-)particles} associated with these modes are generated out of
vacua representing particular solid states.
Exotic exclusion statistics are one manifestation of this diversity.

In the study of magnetic insulators, commonly modelled as lattice systems of interacting spins,
a variety of particles have been identified to describe their statistical mechanical
properties under specific circumstances \cite{Yosi96, Matt06, LMK09}. 
In each case the physical vacuum (ground state) is configured as the pseudo-vacuum
for particles from one or several species representing collective excitations.
In general, these particles scatter off each other elastically or inelastically.
Particles may decay or form bound states.
An exact solution is almost always out of reach.

To the rare exceptions belong the completely integrable spin chain models. 
In these systems, the particles only interact via two-body elastic scattering. 
The kinematic constraints of one dimension combined with the dynamic constraints of a 
factorizing $S$-matrix impose conservation laws that guarantee an exact solution (via 
Bethe ansatz, for example) \cite{KBI93, Suth04, EFG+05}. 
Many integrable spin model systems have one or several Ising limits (of commuting spin operators), 
where some of the key particle properties such as their exclusion statistics are left intact. 

Here we propose a taxonomy of free particles in Ising chains.
We classify particles according to both structure and function. 
We associate structures with particle species and functions with particle categories.
It appears that Ising chains produce particles of a seemingly unlimited
diversity of structures but of only four categories.

It remains to be seen how robust some of the statistical mechanical properties of the particles
identified here are when non-commuting terms are added to the Ising Hamiltonian. 
The insights reported in this study are not limited to magnetism.
Ising systems are applicable to phenomena in other areas of physics and beyond.
The methodology introduced here is adaptable to applications beyond Ising models.

We consider the spin-1 Ising chain with periodic boundary conditions,
\begin{equation}
  \label{eq:1-1}
  \mathcal{H}= \sum_{l=1}^{N}\Big[JS_{l}^{z}S_{l+1}^{z}
  +D(S_{l}^{z})^{2}-hS_l^z\Big],\quad S_{l}^{z}=0,\pm1,
\end{equation}
where the $J$-term denotes a nearest-neighbor coupling, the $D$-term a quadratic potential, 
and the $h$-term the Zeeman energy.
In higher dimensions this Hamiltonian coincides with the thoroughly studied Blume-Emery-Griffiths 
model \cite{BEG71}.
The statistical mechanics of the spin chain (\ref{eq:1-1}) per se is of limited interest and most 
conveniently worked out via the transfer matrix method \cite{KW41}.
We employ this textbook model as a vehicle to develop an idea of much wider scope.

\begin{figure}[t!]
  \centering
  \includegraphics[width=85mm]{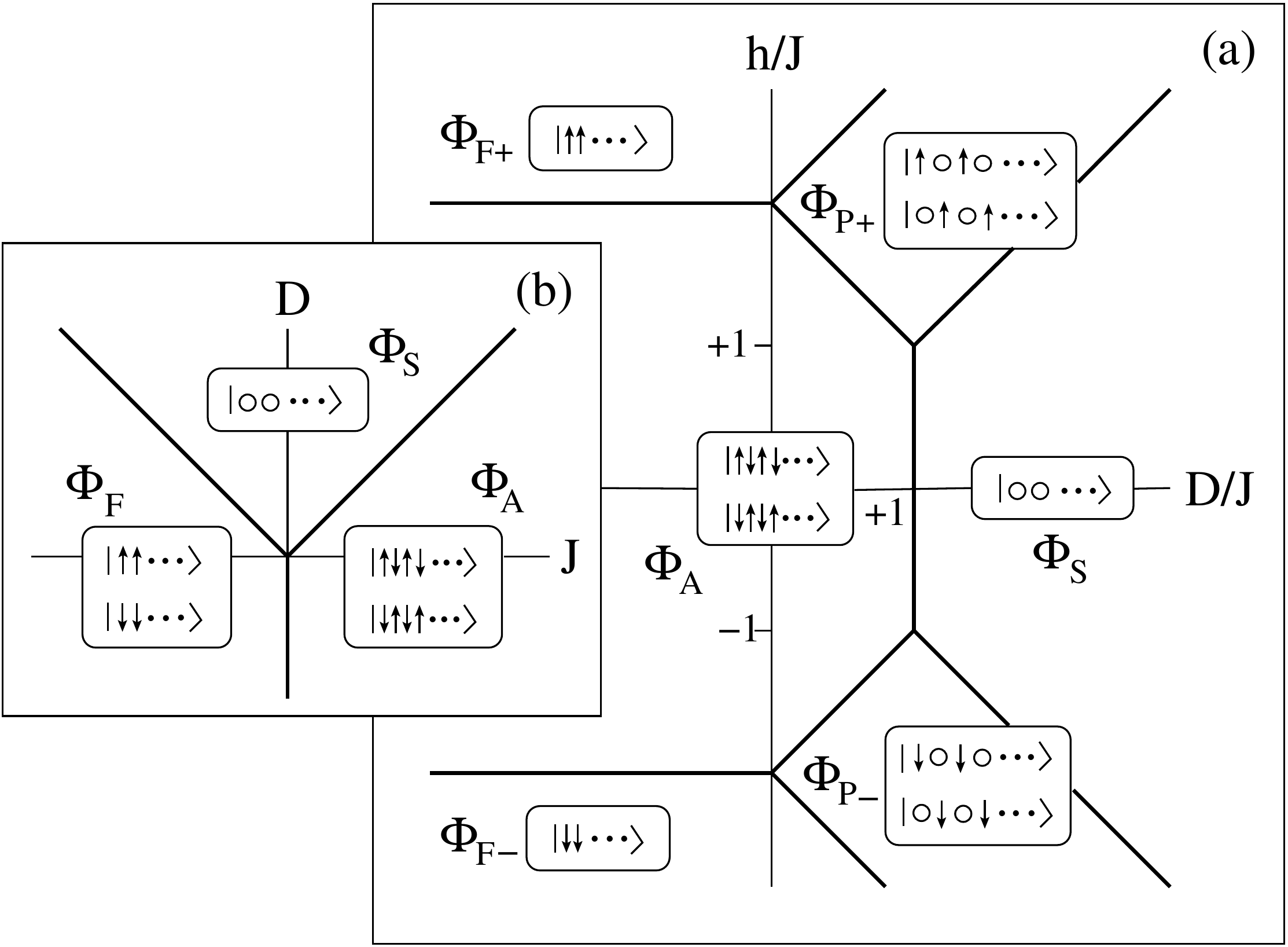} 
  \caption{(a) Six $T=0$ phases $\Phi_{A}$ (antiferromagnetic), $\Phi_{F\pm}$ (ferromagnetic), 
  $\Phi_{S}$ (singlet),   and $\Phi_{P\pm}$ (plateau) for $J>0$; (b) three phases at $h=0$.
  The text refers to regions in parameter space by the same names.} 
  \label{fig:phadah}
\end{figure}

The spectrum of $\mathcal{H}$ is expressible in the form of product states, 
$|\sigma_1\sigma_2\cdots\sigma_N\rangle$ with $\sigma_l=\uparrow,\circ,\downarrow$.
The zero-temperature phase diagram features six phases as shown in 
Fig.~\ref{fig:phadah}(a) \cite{CJS+03}: one antiferromagnetic phase $\Phi_A$, 
one singlet phase $\Phi_S$,
two ferromagnetic phases $\Phi_{F\pm}$, and two plateau phases $\Phi_{P\pm}$ \cite{note1}. 
The physical vacuum (ground state) is unique in $\Phi_{F\pm}$, $\Phi_S$ and twofold degenerate in
$\Phi_A$, $\Phi_{P\pm}$.

In Sec.~\ref{sec:combi} we configure each physical vacuum as the pseudo-vacuum of a set of 
dynamically free but statistically interacting particles.  
The statistical mechanical analysis, carried out in Sec.~\ref{sec:stamec}, yields the exact population 
densities of particles.
In Sec.~\ref{sec:entmix} we interpret the low-$T$ state near phase boundaries 
as that of miscible or immiscible liquids.
Section~\ref{sec:strufunc} further elaborates on the taxonomy of particles.

%
\section{Combinatorial analysis}\label{sec:combi}
%
It will suffice to consider four sets of particles owing to spin-flip symmetry. 
The two sets generated from twofold pseudo-vacua $(\Phi_A,\Phi_{P+})$ have shared attributes.
The two sets generated from unique pseudo-vacua $(\Phi_S,\Phi_{F+})$ share different attributes.
Each choice of pseudo-vacuum can be used in the entire parameter space, but only in one region
does it coincide with the physical vacuum.

Particle species and elements of pseudo-vacuum are characterized by motifs comprising two, 
three or four consecutive site variables $\sigma_l$.
Motifs interlink in one shared site variable.
Particles are subject to mutual statistical interactions as captured in the generalized 
Pauli principle proposed by Haldane \cite{Hald91a}.
The relation,
\begin{equation}\label{eq:2-1} 
\Delta d_m=-\sum_{m'}g_{mm'}\Delta N_{m'},
\end{equation}
expresses how the number of slots available to one particle of species $m$ is affected when particles 
of any species $m'$ from a given set are added.
This idea has proven to be surprisingly versatile with a wide range of applications and a high degree 
of flexibility regarding the notion of particle species \cite{Hald91, Hald94, BW94, LMK09}.

In a chain of $N$ sites the number of many-body states containing a specific combination of 
particles from all species is determined by a multiplicity expression inferred from (\ref{eq:2-1}) via 
combinatorial analysis \cite{Hald91a}.
The exact structure of the multiplicity expression varies between applications.
We have developed a template that can be employed for all situations considered here:
\begin{subequations}\label{eq:2-2} 
\begin{align}
& W(\{N_m\})=\frac{n_{pv}N}{N-N^{(\alpha)}} \prod_{m=1}^M
\left( \begin{tabular}{c}
$d_m+N_m-1$ \\ $N_m$
\end{tabular} \right), \label{eq:2-2a} \\
& d_m=A_m-\sum_{m'=1}^{M}g_{mm'}(N_{m'}-\delta_{mm'}), \label{eq:2-2b}  \\
& N^{(\alpha)}=\sum_{m=1}^M\alpha_mN_m,
\label{eq:2-2c}
\end{align}
\end{subequations}
where $d_m$ counts the number of open slots for particles of species $m$ in the presence of 
$N_{m'}$ particles from any species $m'$.
The permissible combinations $\{N_m\}$ produce integer-valued $d_m$.

Each application is characterized by specific values of $M$ (number of particle species), 
$n_{pv}$ (multiplicity of pseudo-vacuum), $A_m$ (capacity constants), 
$\alpha_m$ (size constants), and $g_{mm'}$ (statistical interaction coefficients). 
These specifications will be tabulated for each situation under scrutiny. 
The prefactor in (\ref{eq:2-2a}) is specific to the use of periodic boundary conditions. 
It is a generalization of a modification factor worked out by Polychronakos \cite{Poly96}.

The energy of a many-body state only depends on its particle content, 
not on the particle configuration:
\begin{equation}\label{eq:2-3} 
E\big(\{N_m\}\big)=E_{pv}+\sum_{m=1}^M N_m\epsilon_m,
\end{equation}
where $E_{pv}$ is the energy of the pseudo-vacuum and $\epsilon_m$ the particle energy 
(species $m$) relative to the pseudo-vacuum. 
There is no interaction energy.
Each phase in Fig.~\ref{fig:phadah}(a) is the pseudo-vacuum of one set of particle species 
or a solid of one species of particles (with $\epsilon_m<0$) from the same set.

All particle motifs occupy at least one bond on the lattice.
Adding particles of any species, therefore, diminishes the number of open slots for more particles.
Normally we have $g_{mm'}>0$, which enforces $\Delta d_m<0$ in (\ref{eq:2-1}) whenever
a particle is added.
The limited space for particles on the finite lattice is signalled in (\ref{eq:2-2a}) by a 
vanishing binomial factor caused by the factorial of a negative $d_m$ in the denominator.

However, we shall find some $g_{mm'}$ to be zero or negative.
A vanishing $g_{mm'}$ means that placing a particle $m'$ has no direct effect on the options for 
placing a particle $m$ thereafter.
A negative $g_{mm'}$ turns the exclusion principle into an accommodation principle.
Placing a particle $m'$ increases the options for placing a particle $m$ thereafter.
In these instances, space limitations on the lattice are enforced by indirect effects on the $d_m$.
The statistical mechanics of a macroscopic system only depends on the specifications
$A_m,\epsilon_m,g_{mm'}$ as will be demonstrated in Sec.~\ref{sec:stamec}.

\subsection{Antiferromagnetic pseudo-vacuum}\label{sec:apv} 
In region $\Phi_{A}$ the ground state is the twofold degenerate N{\'e}el state, 
$|\uparrow\downarrow\uparrow\downarrow\cdots\rangle$, 
$|\downarrow\uparrow\downarrow\uparrow\cdots\rangle$, implying $n_{pv}=2$.
We have identified a set of $M=6$ species of dynamically free particles excited from this 
state \cite{LVP+08}. 
The motifs and statistical interaction specifications are summarized in 
Tables~\ref{tab:specs1neela} and \ref{tab:specs1neelb}.
The full spectrum of product states $|\sigma_1\cdots\sigma_N\rangle$ is equivalent to all 
possible particle configurations.
The number of states with given particle content $\{N_m\}$ and energy (\ref{eq:2-3}) is determined by 
(\ref{eq:2-2a}). 

\begin{table}[h!]
  \caption{Specifications of $M=6$ species of particles excited from the N{\'e}el state $(n_{pv}=2)$, 
  $|\uparrow\downarrow\uparrow\downarrow\cdots\rangle$,
  $|\downarrow\uparrow\downarrow\uparrow\cdots\rangle$: motif, category, species, energy
    (relative to pseudo-vacuum), spin, capacity constants, and size constants. Segments of 
    $\ell$ vacuum bonds $\uparrow\downarrow,\downarrow\uparrow$ have energy 
    $\ell(D-J)$. At $h\neq0$ the entries of $\epsilon_m$ must be amended by 
    $-s_mh$.}\label{tab:specs1neela} 
\begin{center}
\begin{tabular}{ccc|crcc}
motif & category & $m$ & $\epsilon_{m}$ & $s_{m}$ & $A_{m}$ &$\alpha_m$
\\ \hline \rule[-2mm]{0mm}{6mm}
$\uparrow\uparrow$ & compact & $1$ & $2J$ & $+1$ & $\frac{N-1}{2}$ & $1$
\\ \rule[-2mm]{0mm}{4mm}
$\downarrow\downarrow$ & compact & $2$ & $2J$ & $-1$ & $\frac{N-1}{2}$ & $1$
\\ \rule[-2mm]{0mm}{4mm}
$\circ\circ$ & tag & $3$ & $J-D$ & $0$ & $0$ & $1$
\\ \rule[-2mm]{0mm}{4mm}
$\uparrow\circ\uparrow$ & host & $4$ & $2J-D$ & $+1$ & $\frac{N-2}{2}$ & $2$
\\ \rule[-2mm]{0mm}{4mm}
$\downarrow\circ\downarrow$ & host & $5$ & $2J-D$ & $-1$ & $\frac{N-2}{2}$ & $2$
\\ \rule[-2mm]{0mm}{4mm}
$\uparrow\circ\downarrow, \downarrow\circ\uparrow$ & host & $6$ & $2J-D$ & $0$ & $N-1$ & $1$
\end{tabular}
\end{center}
\end{table} 

\begin{table}[t!]
  \caption{Statistical interaction coefficients of the particles excited from the N{\'e}el 
  state (or plateau state.)}\label{tab:specs1neelb} 
\begin{center}
\begin{tabular}{c|rrrrrr}
$g_{mm'}$ & $1$ & $2$ & $3$ & $4$ & $5$ & $6$ \\ \hline \rule[-2mm]{0mm}{6mm}
$1$ & $\frac{1}{2}$ & $\frac{1}{2}$ & $\frac{1}{2}$ & $0$ & $1$ & $\frac{1}{2}$
\\ \rule[-2mm]{0mm}{4mm}
$2$ & $\frac{1}{2}$ & $\frac{1}{2}$ & $\frac{1}{2}$ & $1$ & $0$ & $\frac{1}{2}$
\\ \rule[-2mm]{0mm}{4mm}
$3$ & $~~0$ & $~~0$ & $~~0$ & $-1$ & $-1$ & $-1$
\\ \rule[-2mm]{0mm}{4mm}
$4$ & $\frac{1}{2}$ & $\frac{1}{2}$ & $\frac{1}{2}$ & $1$ & $1$ & $\frac{1}{2}$
\\ \rule[-2mm]{0mm}{4mm}
$5$ & $\frac{1}{2}$ & $\frac{1}{2}$ & $\frac{1}{2}$ & $1$ & $1$ & $\frac{1}{2}$
\\ \rule[-2mm]{0mm}{4mm}
$6$ & $1$ & $1$ & $1$ & $2$ & $2$ & $2$
\end{tabular}
\end{center}
\end{table} 

Particles $m\neq 3$ and vacuum bonds are placed side by side onto the lattice. 
They interlink in one shared site variable as in these examples: 
\begin{align}\label{eq:2-4}
&\uparrow\uparrow\downarrow\circ\downarrow ~\hat{=}~ 1+\mathrm{vac}+5, \qquad 
\downarrow\downarrow\circ\uparrow ~\hat{=}~ 2+6, \nonumber \\
 &\uparrow\circ\uparrow\circ\uparrow ~\hat{=}~ 4+4, \qquad 
\uparrow\circ\uparrow\downarrow\circ\downarrow ~\hat{=}~ 4+\mathrm{vac}+5, \nonumber \\
& \uparrow\circ\downarrow\circ\uparrow ~\hat{=}~ 6+6, \qquad
\uparrow\circ\downarrow\uparrow\circ\downarrow ~\hat{=}~ 6+\mathrm{vac}+6.
\end{align}
Particle 3 can only be placed inside particles 4,5,6 as in the following examples:
\begin{align}\label{eq:2-5}
&\uparrow\circ\circ\uparrow ~\hat{=}~ 4+3, \qquad 
\uparrow\circ\circ\circ\downarrow ~\hat{=}~ 6+3+3, \nonumber \\
&\uparrow\circ\downarrow\circ\circ\downarrow\uparrow\circ\circ\uparrow ~\hat{=}~ 
6+5+3+\mathrm{vac}+4+3.
\end{align}

The six particle species naturally divide into three categories: \emph{compacts}, \emph{hosts}, 
\emph{tags}. 
Particles $m=4,5,6$ are hosts to tag $m=3$. 
The compacts $m=1,2$ do not host nor are they being hosted.

The motifs of compacts 1,2 are to be interpreted as consisting of one bond and half of the 
site at either end. 
In the hosts 4,5,6 we count two bonds and two sites (the site in the middle and 
half of each outer site). 
Tag 3 has one bond and one site. 
The site on the right is counted fully. 
The site on the left is counted as interior site of the surrounding host or as right-hand site 
of another tag to its left.
With these rules each particle has a definite energy $\epsilon_m$ and a definite spin $s_m$. 

Vanishing and negative $g_{mm'}$ are indeed realized in Table~\ref{tab:specs1neelb}.
The placement of one host, $m'=4$, does not affect the number of open slots for 
placing a compact, $m=1$; hence $g_{14}=0$.
Likewise, the options for placing a tag, $m=3$, are unaffected by first placing 
a compact $m'=1,2$, or another tag; hence $g_{31}=g_{32}=g_{33}=0$.

If a host $m'=4,5,6$ is placed into any open slot then the options for placing
a tag $m=3$ increases by one; hence $g_{34}=g_{35}=g_{36}=-1$.
Tags have vanishing capacity constants: $A_3=0$.
Open slots for tags are created by hosts.

The two distinct motifs of host 6 do not compromise the indistinguishability of particles 
belonging to the same species.
At most one of the two motifs can be placed into any particular slot. 
Distinguishable motifs of indistinguishable particles cannot exchange positions. 
It is possible to split species 6 into two species with one motif each. 
This introduces a mutual statistical interaction between the two new species and modifies  the 
statistical interactions with all other species \cite{Angh07, note7}.

In region $\Phi_A$ all $\epsilon_m$ are positive.
Here the physical vacuum is the pseudo-vacuum.
Elsewhere in parameter space we have $\epsilon_m<0$ for at least one species.
There the physical vacuum is a solid formed by the particles with the lowest (negative) energy density.
We can reconfigure any one of these solids as the pseudo-vacuum for a different set of particles.
In Ref.~\cite{LVP+08} these particles were interpreted as soliton pairs in generalization of the 
well-known classification used in the context of spin-$\frac{1}{2}$ $XXZ$ chains \cite{VIS}.

\subsection{Singlet pseudo-vacuum}\label{sec:spv} 
In region $\Phi_{S}$ the ground state is unique, $|\circ\circ\cdots\circ\rangle$, implying 
$n_{pv}=1$. 
It is the pseudo-vacuum for a set of $M=7$ particle species.
The specifications are listed in Tables~\ref{tab:specs1singleta} and \ref{tab:specs1singletb}.
The multiplicity expression (\ref{eq:2-2}) again accounts for the full spectrum.

\begin{table}[htb]
  \caption{Specifications of $M=7$ particles excited from the singlet state $(n_{pv}=1)$, 
    $|\circ\circ\cdots\rangle$: motif, category, species, energy (relative to pseudo-vacuum), 
    spin, capacity constants, and size constants. 
    Segments of vacuum bonds $\circ\circ$ have zero energy. At $h\neq0$ the entries of $\epsilon_m$ 
    must be amended by $-s_mh$.}\label{tab:specs1singleta} 
\begin{center}
\begin{tabular}{ccc|crcc}
motif & category & $m$ & $\epsilon_{m}$ & $s_{m}$ & $A_{m}$ & $\alpha_m$
\\ \hline 
$\circ\!\uparrow\!\circ$ & host & $1$ & $D$ & $+1$ & $N-1$ & $1$
\\ 
$\circ\!\downarrow\!\circ$ & host &$2$ & $D$ & $-1$ & $N-1$ & $1$
\\ 
$\circ\!\uparrow\downarrow\!\circ$ & host &$3$ & $2D-J$ & $0$ & $N-2$ & $2$
\\ 
$\circ\!\downarrow\uparrow\!\circ$ & host & $4$ & $2D-J$ & $0$ & $N-2$ & $2$
\\ 
$\uparrow\uparrow$ & hybrid & $5$ & $D+J$ & $+1$ & $0$ & $1$
\\ 
$\downarrow\downarrow$ & hybrid & $6$ & $D+J$ & $-1$ & $0$ & $1$
\\ 
$\uparrow\downarrow\uparrow, \downarrow\uparrow\downarrow$ & hybrid & $7$ 
& $2D-2J$ & $0$ & $0$  & $2$ 
\end{tabular}
\end{center}
\end{table} 

\begin{table}[htb]
  \caption{Statistical interaction coefficients of the particles excited from the singlet 
  state (or spin-polarized state.)}\label{tab:specs1singletb} 
\begin{center}
\begin{tabular}{c|rrrrrrr}
$g_{mm'}$ & $1$ & $2$ & $3$ & $4$ & $5$ & $6$ & $7$
\\ \hline \rule[-2mm]{0mm}{6mm}
$1$ & $2$ & $1$ & $2$ & $2$ & $1$ & $1$ & $2$ \\ 
$2$ & $2$ & $2$ & $2$ & $2$ & $1$ & $1$ & $2$ \\ 
$3$ & $2$ & $2$ & $3$ & $2$ & $1$ & $1$ & $2$ \\ 
$4$ & $2$ & $2$ & $3$ & $3$ & $1$ & $1$ & $2$ \\ 
$5$ & $-1$ & $0$ & $-1$ & $-1$ & $~~0$ & $~~0$ & $-1$ \\ 
$6$ & $0$ & $-1$ & $-1$ & $-1$ & $0$ & $0$ & $-1$ \\ 
$7$ & $-1$ & $-1$ & $-1$ & $-1$ & $0$ & $0$ & $0$ \\
\end{tabular}
\end{center}
\end{table} 

There are no compacts among the seven species. 
Particles 1,2,3,4 are hosts to particles 5,6,7.
The latter behave like tags in some respects.
However, their ability to coexist inside the same host implies a hosting capability among them.
We have named them \emph{hybrids}.
Particle 7 has two motifs, which again causes no complications.

Each host is an island in the singlet pseudo-vacuum.
The size of an island depends on its hybrid content.
When a hybrid is placed inside a host the first site variable (on the left) of its motif is shared
with an interior site variable of the host or any matching site variable of a hybrid already
present inside that host as in these examples:
\begin{align}\label{eq:2-6} 
& \circ\uparrow\uparrow\circ ~\hat{=}~ 1+5,\qquad 
\circ\uparrow\downarrow\uparrow\downarrow\circ ~\hat{=}~ 3+7, \nonumber \\
& \{\circ\uparrow\downarrow\uparrow\uparrow\circ, \circ\uparrow\uparrow\downarrow\uparrow\circ\}
~\hat{=}~ 1+5+7.
\end{align}
In the last instance, two states correspond to the same combinations of particles, which is accounted
for by the multiplicity expression (\ref{eq:2-2}).

In the course of adding a hybrid, one bond of the host or of a hybrid already present is broken up 
and replaced by an identical bond to the right of the newly placed hybrid.
Each host has at most one open slot for any particular hybrid.
The second example in (\ref{eq:2-6}) is no exception because it makes no difference whether one 
or the other motif of hybrid 7 is placed inside host 3. 

Whereas a given combination of particles may be placed in multiple ways on the lattice, each product 
state has a unique particle content as expressed by the list $\{N_m\}$ in 
(\ref{eq:2-2a}) and (\ref{eq:2-3}).
The rules that ensure that each bond and each site are counted exactly once  are the same as
in Sec.~\ref{sec:apv}.

The $g_{mm'}$ listed in Table~\ref{tab:specs1singletb} again include some vanishing or negative
coefficients. 
Not all occurrences have the same interpretation.
Hosts $m'=3,4$ can accommodate all hybrids $m=5,6,7$ whereas hybrid $m=5~(6)$ cannot be 
accommodated by host $m'=2~(1)$ without the presence of at least one hybrid $m=7$.
This explains the non-positive coefficients $g_{mm'}$ between hosts and hybrids.

Ambiguities arise when we aim to determine the $g_{mm'}$ among hybrids. 
Hybrids $m=5,6$ alone do not share open slots for placement, implying $g_{56}=g_{65}=0$.
New situations present themselves when hybrids $m=7$ are added to the mix.
All hosts are capable of accommodating hybrids $m=5,7$ or $m=6,7$ or $m=5,6,7$ 
in multiple combinations. 

Consider a single host, say $m=1$, with several hybrids embedded. 
If all hybrids are of the same species, e.g. $m'=5$ or $m'=7$, then the multiplicity is dictated 
by the size of the expanded host. 
This information is encoded in the $g_{mm'}$ between host $m$ and hybrid $m'$. 

However, if host $m=1$ accommodates hybrids $m'=5,7$ in several different sequences, then that 
contribution to the multiplicity is encoded in the statistical interaction coefficients between the 
two hybrids. 
Since there is no hierarchy between the two hybrids, the encoding of that information can be 
implemented in $g_{75}$ or in $g_{57}$. 
In Table~\ref{tab:specs1singletb} we use $g_{57}=g_{67}=-1$, $g_{75}=g_{76}=0$.  
Different choices produce the same multiplicity.
The hybrids (like the tag in Sec.~\ref{sec:apv}) have zero capacity constants $A_m$.

\subsection{Ferromagnetic pseudo-vacuum}\label{sec:fpv} 
Here we pick the spin-polarized state, $|\uparrow\uparrow\cdots\rangle$, as the pseudo-vacuum
for a set of particle species that again generate the full spectrum.
This state coincides with the physical vacuum in region $\Phi_{F+}$.
The specifications for a set of $M=7$ particles can be obtained, in this instance, by reasoning as 
follows. 
With the transcription 
\begin{equation}\label{eq:2-7} 
\{\circ,\uparrow,\downarrow\}_S ~\hat{=}~ \{\uparrow,\circ,\downarrow\}_{F+}
\end{equation}
between site variables we produce a one-on-one mapping between Ising product states that  maps 
the singlet state $|\circ\circ\cdots\rangle$ onto the spin-polarized state 
$|\uparrow\uparrow\cdots\rangle$. 
Given that the singlet state is the pseudo-vacuum of the particles listed in 
Table~\ref{tab:specs1singleta}, the transcription yields a corresponding set of particles 
generated from the spin-polarized state. 
These particles are listed in Table~\ref{tab:specs1spipoa}. 

Hosts map onto hosts and hybrids onto hybrids, leaving the $A_m$ and $\alpha_m$ invariant.
The $g_{mm'}$ keep the values listed in Table~\ref{tab:specs1singletb}.
The mapping changes the specifications $\epsilon_m,s_m$.
Given the nonzero spin-polarization of the pseudo-vacuum, the quantum number $s_m$
is not a spin in the usual sense, but its use enables us to write the magnetic-field dependence of the 
particle energy in the Zeeman form, $-s_mh$.

\begin{table}[t!]
  \caption{Specifications of $M=7$ particles excited from the
    spin-polarized state $(n_{pv}=1)$, ${|\uparrow\uparrow\cdots\rangle}$: motif,
    category, species, energy (relative to pseudo-vacuum), `spin', capacity constants, and size constants. 
    Segments of $\ell$ vacuum bonds $\uparrow\uparrow$ have energy $\ell(D+J)$. At $h\neq0$ the 
    entries of $\epsilon_m$ must be amended by $-s_mh$.}\label{tab:specs1spipoa} 
\begin{center}
\begin{tabular}{ccc|cccc}
motif & category & $m$ & $\epsilon_{m}$ & $s_m$ & $A_{m}$ & $\alpha_m$
\\ \hline 
 $\uparrow\!\circ\!\uparrow$ & host &$1$ & $-2J-D$ & $-1$ & $N-1$ & $1$
\\ 
$\uparrow\downarrow\uparrow$ & host & $2$ & $-4J$ & $-2$ & $N-1$ & $1$ 
\\ 
$\uparrow\!\circ\!\downarrow\uparrow$ & host & $3$ & $-4J-D$ & $-3$ & $N-2$ & $2$ 
\\ 
$\uparrow\downarrow\!\circ\!\uparrow$ & host & $4$ & $-4J-D$ & $-3$ & $N-2$ & $2$ 
\\ 
$\circ\circ$ & hybrid & $5$ & $-J-D$ & $-1$ & $0$ & $1$ 
\\ 
$\downarrow\downarrow$ & hybrid & $6$ & $0$ & $-2$ & $0$ & $1$ 
\\ 
$\circ\!\downarrow\!\circ, \downarrow\!\circ\!\downarrow$ & hybrid & $7$ & 
$-2J-D$ & $-3$ & $0$  & $2$
\end{tabular}
\end{center}
\end{table} 

\subsection{Plateau pseudo-vacuum}\label{sec:ppv} 
The partially magnetized plateau state, $|\uparrow\circ\uparrow\circ\cdots\rangle$, 
$|\circ\uparrow\circ\uparrow\cdots\rangle$, which is the physical vacuum in region $\Phi_{P+}$,
is selected here as the pseudo-vacuum for the fourth set of particle species.
These particles can be found with the transcription
\begin{equation}\label{eq:2-8} 
\{\circ,\uparrow,\downarrow\}_A ~\hat{=}~ \{\downarrow,\downarrow,\circ\}_{P+}
\end{equation}
between site variables, which produces a one-on-one mapping between the N\'eel state and 
the plateau state. 
The specifications of the $M=6$ particles transcribed from Table~\ref{tab:specs1neela} are listed 
in Table~\ref{tab:specs1plat7a}. 
Only the motifs and the entries of $\epsilon_m,s_m$ are different.
The $g_{mm'}$ remain as stated in Table~\ref{tab:specs1neelb}.
The transcription does not affect the categories.

\begin{table}[htb]
  \caption{Specifications of $M=6$ particles excited from the     
  plateau state $(n_{pv}=2)$, $|\uparrow\circ\uparrow\circ\cdots\rangle$,
  $|\circ\uparrow\circ\uparrow\cdots\rangle$: motif, category, species, energy
    (relative to pseudo-vacuum), `spin', capacity constants, and size constants. 
    Segments of $\ell$ vacuum     bonds $\uparrow\circ,\circ\uparrow$ have energy 
    $\ell D/2$. At $h\neq0$ the entries for $\epsilon_m$ must be amended by $-
    s_mh$.}\label{tab:specs1plat7a} 
\begin{center}
\begin{tabular}{ccc|cccc}
motif & category & $m$ & $\epsilon_{m}$ & $s_m$ & $A_{m}$ & $\alpha_m$
\\ \hline \rule[-2mm]{0mm}{6mm}
$\uparrow\uparrow$ & compact & $1$ & $J+\frac{1}{2}D$ & $~~\frac{1}{2}$ & $\frac{N-1}{2}$ & $1$
\\ \rule[-2mm]{0mm}{4mm}
$\circ\circ$ & compact & $2$ & $-\frac{1}{2}D$ & $-\frac{1}{2}$ & $\frac{N-1}{2}$ & $1$
\\ \rule[-2mm]{0mm}{4mm}
$\downarrow\downarrow$ & tag & $3$ & $J+\frac{1}{2}D$ & $-\frac{3}{2}$ & $0$ & $1$
\\ \rule[-2mm]{0mm}{4mm}
$\uparrow\downarrow\uparrow$ & host & $4$ & $-2J+D$ & $-1$ & $\frac{N-2}{2}$ & $2$
\\ \rule[-2mm]{0mm}{4mm}
$\circ\downarrow\circ$ & host & $5$ & $0$ & $-2$ & $\frac{N-2}{2}$ & $2$
\\ \rule[-2mm]{0mm}{4mm}
$\uparrow\downarrow\circ,\circ\downarrow\uparrow$  & host & $6$ & $-J+\frac{1}{2}D$ & 
$-\frac{3}{2}$ & $N-1$ & $1$
\end{tabular}
\end{center}
\end{table} 

The structural correspondence between particle species as determined by transcriptions 
(\ref{eq:2-7}) and (\ref{eq:2-8}) does not imply a functional correspondence because
some functions depend on the relative energies as will be discussed in 
Secs.~\ref{sec:entmix} and \ref{sec:strufunc}.

\subsection{Species and categories}\label{sec:specat} 
The following comments are based on wider evidence than reported in 
this work, including evidence from Ising chains with spin $s=\frac{1}{2},\frac{3}{2}$ and with 
next-nearest-neighbor coupling, where sets of particles with up to 17 species and motifs of up to 
six sites have been found \cite{note3}. 

(i) The template (\ref{eq:2-2}) for the multiplicity expression with specs 
$n_{pv}, A_m, \alpha_m, g_{mm'}$ characterizing the pseudo-vacuum and all particle species 
excited from it has proven sufficiently robust to hold for all cases investigated.

(ii) The classification of particles into the four categories of compacts, hosts, tags, 
and hybrids has proven useful and comprehensive within the range of our explorations. 
Compacts and hosts find open slots on segments of pseudo-vacuum, tags and hybrids inside hosts, 
hybrids with hosting capability for different hybrids.

(iii) The capacity constants $A_m$ are extensive $(\propto N)$ for compacts and hosts.
The value of $A_m$ reflects the holding capacity of the pseudo-vacuum for particles of these 
two categories. 
It depends on the size of the motif(s) of a given species and how closely particles of that species 
can be packed.
The pseudo-vacuum has no holding capacity for tags and hybrids, implying $A_m=0$.
Holding capacity for tags and hybrids must be generated by hosts.

(iv) The statistical interaction coefficient $g_{mm'}$ reflects the direct impact on the open slots
available to particles of species $m$ if one particle of species $m'$ is added. 
Negative coefficients $g_{mm'}$ represent an accommodation principle.
The presence of hosts $m'$ opens up slots for tags or hybrids $m$.
However, for every $g_{mm'}<0$ there is a $g_{m'm}>0$, which, in combination, limit the holding 
capacity for hosts and tags (or hybrids).

(v) The limited room for particles of any species is signalled in the multiplicity
expression (\ref{eq:2-2}) by a vanishing binomial factor as caused by factorial of a negative $d_m$
(interpreted as a $\Gamma$-function) in the denominator. 

%
\section{Statistical mechanical analysis}\label{sec:stamec}
%
The statistical mechanics of the four sets of particles identified in Sec.~\ref{sec:combi}
is amenable to a rigorous analysis that starts from the multiplicity expression (\ref{eq:2-2}).
The method was developed by Wu \cite{Wu94} for a generic situation 
(see also Refs.~\cite{Isak94, NW94}).

The range of applications is surprisingly wide.
It includes an alternative to the thermodynamic Bethe ansatz \cite{BW94}, applications to 
particles in real space and in reciprocal space \cite{LMK09, Hald91a, Hald94, FS98, KMW08}, 
and applications to systems in higher dimensions \cite{PMK07}.
Each application requires a certain degree of adaptation of Wu's generic result.
The application to interlinking particles with shapes is no exception.

For given energies $\epsilon_m$ and statistical interaction specifications $A_m$, $g_{mm'}$,
the grand partition function of any one of the four sets of particles has the form
\begin{equation}\label{eq:3-1} 
Z=\prod_{m=1}^M\left(\frac{1+w_m}{w_m}\right)^{A_m},
\end{equation}
where the (real, positive) $w_m$ are the solutions of the coupled nonlinear algebraic equations,
\begin{equation}\label{eq:3-2} 
\frac{\epsilon_m}{k_BT}=\ln(1+w_m)-\sum_{m'=1}^Mg_{m'm}\ln\left(\frac{1+w_{m'}}{w_{m'}}\right).
\end{equation}
With the $w_m$ thus determined, the average number of particles, $\langle N_m\rangle$, 
are the solutions of the coupled linear equations \cite{note2},
\begin{equation}\label{eq:3-3} 
w_m\langle N_m\rangle+\sum_{m'=1}^Mg_{mm'}\langle N_{m'}\rangle =A_m.
\end{equation}

Note the sensitivity of Eqs.~(\ref{eq:3-1}) and (\ref{eq:3-3}) to particle category.
Tags and hybrids, which have $A_m=0$, do not contribute their own factor to $Z$.
Their contributions manifest themselves indirectly via their hosts.
Equations~(\ref{eq:3-3}) with indices $m$ pertaining to tags or hybrids
are homogeneous, which makes the tag and hybrid populations conditional on host populations.

The solution of Eqs.~(\ref{eq:3-1})-(\ref{eq:3-3}) covers a wider territory than the transfer-matrix 
solution of the Ising chain (\ref{eq:1-1}) because we are free to assign any values to the
particles energies $\epsilon_m$ \cite{note10}.
We choose the tabulated values pertaining to the three-parameter Ising model for mere convenience.
The ordering tendencies and phase boundaries are then as shown in Fig.~\ref{fig:phadah}.

In the following we determine the average population densities,
$\langle \bar{N}_m\rangle\doteq\langle N_m\rangle/N$, of particles from all species. 
These quantities inform us about the roles of particle species in the ordering tendencies 
and about the significance of particle categories in the behavior near phase boundaries.

\subsection{View from $\Phi_A$}\label{sec:smva}  
The statistical mechanical analysis of the particles identified in Sec.~\ref{sec:apv}
begins with the solution of Eqs.~(\ref{eq:3-2}).
The excitation energies $\epsilon_{m}$ including Zeeman term, $-s_mh$, are taken from 
Table~\ref{tab:specs1neela} and the statistical interaction coefficients from Table~\ref{tab:specs1neelb}. 
The six equations with these specifications can be simplified into
\begin{align}\label{eq:3-5} 
& \frac{w_6}{w_3}=e^{K_J} ,\quad \frac{1+w_2}{1+w_3}=e^{K_J+K_D+H},\quad
\frac{1+w_2}{1+w_1}=e^{2H}, \nonumber \\
& \frac{w_1w_3}{1+w_4}=e^{K_J},\quad \frac{w_2}{w_1}\frac{1+w_4}{1+w_5}=1,\quad 
\frac{\sqrt{w_4w_5}}{1+w_6}=1,
\end{align}
where $K_J\doteq J/k_BT$, $K_D\doteq D/k_BT$, $H\doteq h/k_BT$ are scaled parameters.
Expression (\ref{eq:3-1}) for the grand partition function becomes
\begin{equation}\label{eq:3-6} 
Z=\Big[(1+w_3)e^{K_D-K_J}\Big]^N.
\end{equation}
Equations (\ref{eq:3-5}) reduce to the cubic equation for $w_3$ \cite{note4}. The solution at $h=0$ is 
sufficiently compact to be cited:
\begin{align}\label{eq:3-6-1} 
1+w_3 &= e^{-K_D}\cosh K_J+\frac{1}{2} \nonumber \\
&+\sqrt{\left(e^{-K_D}\cosh K_J-\frac{1}{2}\right)^2+2e^{-K_D}}.
\end{align}

We calculate the dependence of the population densities $\langle\bar{N}_m\rangle$ on the reduced 
parameters $D/J$ and $h/J$ at a fixed value of reduced temperature $k_BT/J$ by solving 
Eqs.~(\ref{eq:3-3}) for given $w_m$. 

In the high-$T$ limit we have $w_1=w_2=w_3=w_6=2$, $w_4=w_5=3$, and
$\langle\bar{N}_1\rangle=\langle\bar{N}_2\rangle=\langle\bar{N}_3\rangle=
\langle\bar{N}_6\rangle=\frac{1}{9}$, $\langle\bar{N}_4\rangle=\langle\bar{N}_5\rangle=\frac{1}{18}$,
implying $Z=3^N$, the total multiplicity of states.

Taking into account the sizes and multiplicity of motifs the population densities must satisfy 
the constraint,
\begin{equation}\label{eq:3-6-5} 
\langle\bar{N}_1\rangle+\langle\bar{N}_2\rangle+\langle\bar{N}_3\rangle+
2\Big(\langle\bar{N}_4\rangle+\langle\bar{N}_5\rangle+\langle\bar{N}_6\rangle\Big)
+\langle\bar{N}_\mathrm{vac}\rangle=1, 
\end{equation}
where $\langle\bar{N}_\mathrm{vac}\rangle$ is the density of vacuum elements with motifs
$\uparrow\downarrow,\downarrow\uparrow$.
Relation~(\ref{eq:3-6-5}) holds at any $T$. 
For $T=\infty$ we infer $\langle\bar{N}_\mathrm{vac}\rangle=\frac{2}{9}$.

\begin{figure}[t]
\includegraphics[width=85mm]{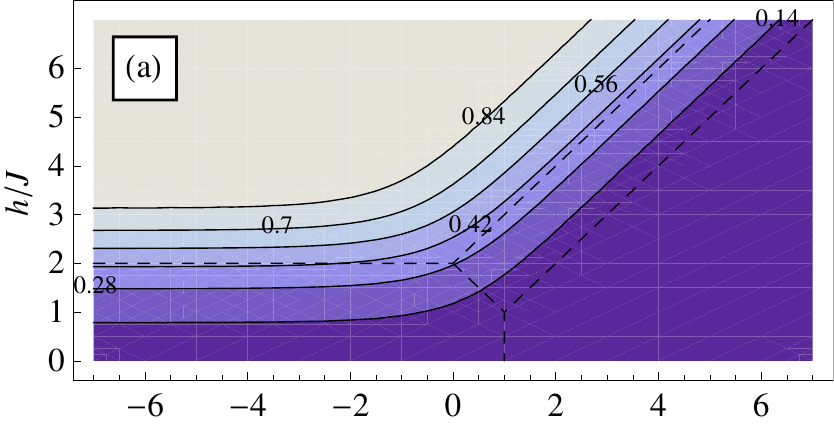}
\includegraphics[width=85mm]{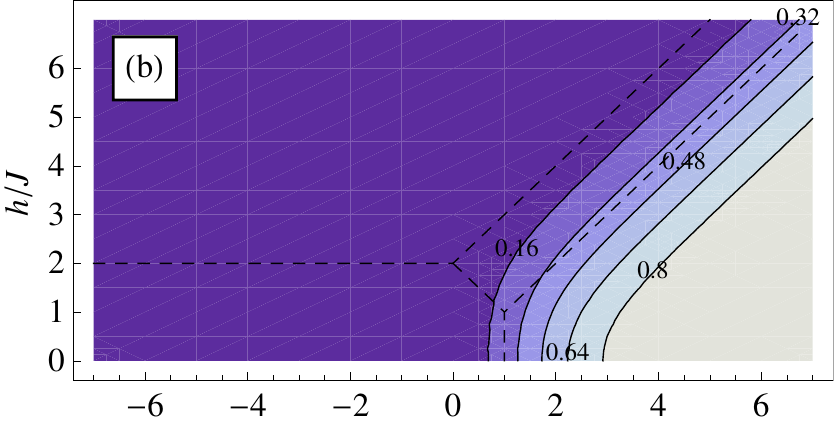}
\includegraphics[width=85mm]{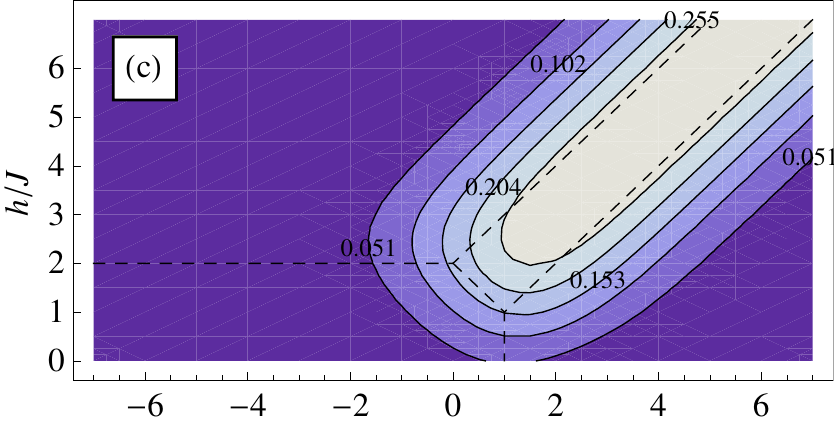}
\includegraphics[width=85mm]{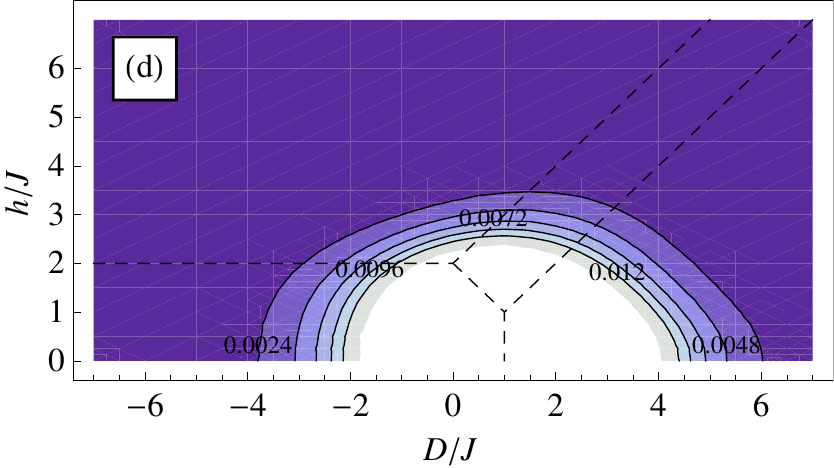}
  \caption{Average numbers (per site) $\langle \bar{N}_{m}\rangle$
  of (a) compacts $m=1$, (b) tags $m=3$,  (c) hosts $m=4$, and (d) hosts $m=6$ versus $D/J$ and 
  $h/J$ at $k_BT/J=1$. The dashed lines indicate the phase boundaries as in 
  Fig.~\ref{fig:phadah}(a)}\label{fig:dlj2-phiah1}
\end{figure}

As $T$ is lowered toward zero, particles from all species with $\epsilon_m>0$ are gradually frozen out 
and particles from some species with $\epsilon_m<0$ are crowded out by particles from the species with 
the lowest negative energy per bond.
Inside any region in the parameter space of Fig.~\ref{fig:phadah}(a) the state of the system in the limit 
$T\to0$ (physical vacuum) has either been emptied out of any particles (pseudo-vacuum) or has turned 
into a solid of one species of particles.
The former is the case in region $\Phi_A$, the latter in all other regions \cite{note8}.

Compacts $m=1$ have negative energies $(\epsilon_1<0)$ in region $\Phi_{F+}$. 
Hence their high density in this region as is evident in Fig.~\ref{fig:dlj2-phiah1}(a). 
Phase $\Phi_{F+}$ is a solid of interlinked compacts 1. 
Tags 3 have negative energies in the region $\Phi_S$. 
That is reflected in Fig.~\ref{fig:dlj2-phiah1}(b). 
These particles solidify at $T=0$ in that region. 

The plateau phase $\Phi_{P+}$ is a solid of interlinked hosts 4 (with $\epsilon_4<0$). 
The high density of these hosts in the narrow region $\Phi_{P+}$ at nonzero $T$ is 
illustrated in Fig.~\ref{fig:dlj2-phiah1}(c). 
Hosts 6 have positive energy for any value of $h/J$ and $D/J$. 
Their average number remains small everywhere but reaches a flat maximum at $h/J\simeq D/J\simeq0$ 
as is illustrated in Fig.~\ref{fig:dlj2-phiah1}(d). 

The plots for $\langle\bar{N}_3\rangle$ and $\langle\bar{N}_6\rangle$ are reflection symmetric
with respect to the line $h/J=0$.
Plots for $\langle\bar{N}_2\rangle$ and $\langle\bar{N}_5\rangle$ would be mirror images of the 
plots for $\langle\bar{N}_1\rangle$ and $\langle\bar{N}_4\rangle$, respectively.

At the five phase boundaries that border region $\Phi_A$ we have $\epsilon_m=0$ for exactly one 
species $m$.
In all but one case that species is a compact or a host.
Particles from these categories interlink directly with elements of pseudo-vacuum and thus produce
a state of high degeneracy.

The $\Phi_A-\Phi_S$ phase boundary is the exception.
Here the particles with $\epsilon_m=0$ are tags.
They do not interlink with elements of pseudo-vacuum.
They exist only inside hosts but all hosts have $\epsilon_m>0$.
The tags $m=3$ by themselves can only form one zero-energy state, the singlet singlet state 
$|\circ\circ\cdots\rangle$.
The degeneracy remains low (threefold).

At the four phase boundaries not bordering region $\Phi_A$ two species of particles have equal 
(negative) energy per bond.
In two cases $(\Phi_S-\Phi_{P\pm})$ the two species are a host and a tag, either $m=4,3$ or $m=5,3$.
In the other two cases $(\Phi_{P\pm}-\Phi_{F\pm})$ the two species are a compact and a host, 
either $m=1,4$ or $m=2,5$.
In all these cases the particle species involved interlink either side by side or one inside 
the other in many different configurations. 

\subsection{View from $\Phi_S$}\label{sec:smvs}  
\begin{figure}[t]
  \includegraphics[width=85mm]{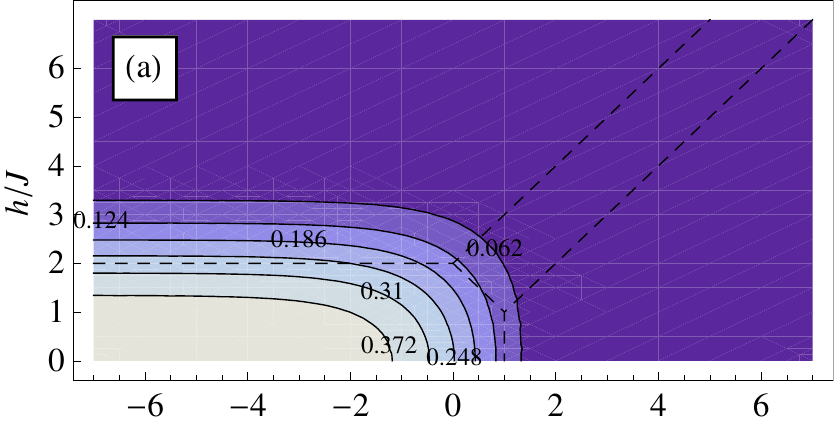}
  \includegraphics[width=85mm]{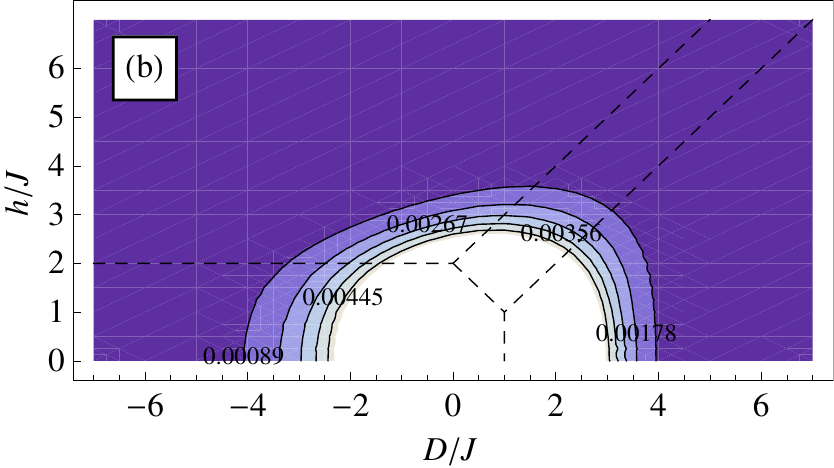}
\caption{Average numbers (per site) $\langle \bar{N}_{m}\rangle$ 
 of (a) hybrids $m=7$ and (b) hosts $m=3,4$ versus $D/J$ and $h/J$ at $k_BT/J=1$.}
  \label{fig:dlj2-phish}
\end{figure}

The analysis for the particles identified in Sec.~\ref{sec:spv} proceeds along the same lines. 
Equations~(\ref{eq:3-2}) for this set of particles can be simplified into
\begin{align}\label{eq:3-7} 
& \frac{1+w_6}{1+w_5}=e^{2H},\quad \frac{w_5w_6}{1+w_7}=e^{4K_J},\quad 
\frac{w_4}{w_7}=e^{K_J}, \nonumber \\
& \frac{(1+w_1)w_6}{(1+w_6)w_7}=e^{2K_J-K_D-H}, \quad
\frac{w_3}{1+w_4}=1, \nonumber \\
&\frac{w_1w_6}{(1+w_2)w_5}=1,\quad
\frac{w_1w_2w_6}{(1+w_2)(1+w_3)}=e^{2K_{J}}. 
\end{align}
They reduce to a cubic equation for $w_5$ (quadratic for $h=0$).
Expression (\ref{eq:3-1}) for the grand partition function reduces to \cite{note5}
\begin{equation}\label{eq:43-3} 
Z=\Big[(1+w_5)e^{-K_J-K_D+H}\Big]^N.
\end{equation}
The dependence of the $\langle \bar{N}_m\rangle$ on $D/J$ and $h/J$ at fixed $k_BT/J$ 
as inferred from Eqs.~(\ref{eq:3-3}) are (selectively) depicted in Fig.~\ref{fig:dlj2-phish}.

At $T=\infty$ we have the values
$w_1=\frac{7}{2}$,$w_2=\frac{5}{2}$,$ w_3=4$, $w_4=3$, $w_5=w_6=2$, $w_7=3$; 
$\langle\bar{N}_1\rangle=\langle\bar{N}_2\rangle= \langle\bar{N}_7\rangle =\frac{2}{27}$, 
$\langle\bar{N}_3\rangle=\langle\bar{N}_4\rangle=\frac{1}{27}$, 
$\langle\bar{N}_5\rangle=\langle\bar{N}_6\rangle=\frac{1}{9}$.
At finite $T$ the population densities are low for all particles in region $\Phi_S$. 
Here all particles have positive energy, the ground state is free of particles, the physical vacuum 
is the pseudo-vacuum. 
In Sec.~\ref{sec:smva} the same physical vacuum was a solid of tags from a different set of particles.

Hybrids 7 have negative energy in region $\Phi_A$,
causing a high population density there as is evident in Fig.~\ref{fig:dlj2-phish}(a).
Phase $\Phi_A$, which was the pseudo-vacuum of choice in Sec.~\ref{sec:smva}, now is a solid formed 
by these hybrids. 

In all other regions the ground state is a solid of one species of particles from both views.
In regions $\Phi_{F\pm}$ the particles with dominating population densities at low $T$ are now hybrids 
5 or 6. 
The contour plots  (not shown) look very similar to Fig.~\ref{fig:dlj2-phiah1}(a) and its mirror image. 

Hosts 1 and 2 have a high population in regions $\Phi_{P\pm}$ and solidify into plateau phases 
as $T\to0$. 
The contour plots for $\langle \bar{N}_1\rangle$, $\langle \bar{N}_2\rangle$ look very similar to 
Fig.~\ref{fig:dlj2-phiah1}(c) and its mirror image.
Hosts 3 and 4 tend to be frozen out or crowded out by other particle species in all regions.
Their population densities remain low as shown in Fig.~\ref{fig:dlj2-phish}(b). 

At two of the three phase boundaries that border region $\Phi_S$ the particle species with 
$\epsilon_m=0$ are hosts $(m=1,2)$, which interlink with elements of vacuum, and at the third phase 
boundary they are hybrids $(m=7)$, which do not. 
This explains the high degeneracy at the $\Phi_S-\Phi_{P\pm}$ boundaries and the low degeneracy at the 
$\Phi_S-\Phi_A$ boundary from the vantage point of the singlet pseudo-vacuum.

At the remaining phase boundaries we again have two species of particles with equal and lowest 
(negative) energy per bond.
In all cases particles of the two species interlink and thus produce a high degeneracy albeit not for the 
same reason.
In four cases they are hosts and hybrids.
In two cases they are two species of hybrids.
 
\subsection{Views from $\Phi_{F\pm}$ and $\Phi_{P\pm}$}\label{sec:smvf}  
Equations~(\ref{eq:3-2}) for the particles excited from the spin-polarized state (Sec.~\ref{sec:fpv})
can be simplified into Eqs.~(\ref{eq:3-7}) with the right-hand sides changed into
\begin{align}\label{eq:3-8} 
& e^{K_J+K_D+H},\quad e^{K_J},\quad e^{-2K_J},\nonumber \\
&  e^{-2H},\quad  1,\quad e^{3K_J},\quad e^{2K_J}.
\end{align}
The close relationship originates in the transcription (\ref{eq:2-7}) between the motifs of the two sets 
of particles.
The grand partition function for this case reads
\begin{equation}\label{eq:3-9} 
Z=\Big[(1+w_6)e^{-2H}\Big]^N.
\end{equation}

The results for the $\langle \bar{N}_m\rangle$  as inferred from Eqs.~(\ref{eq:3-3}) confirm what is 
readily predicted by analogy to the previous cases. 
At low $T$ we have a low density of any particle species in region $\Phi_{F+}$ because the physical 
vacuum of this region is the pseudo-vacuum.
In regions $\Phi_{P+}, \Phi_A$ we find high densities of hosts $m=1,2$, respectively, and in regions
$\Phi_S, \Phi_{F-}, \Phi_{P-}$ high densities of hybrids $m=5,6,7$, respectively. 
Hosts $m=3,4$ do not reach significant densities in any region.

At all phase boundaries except one the particles present in high density interlink directly among 
themselves or with elements of vacuum, thus producing a high degeneracy.
The low-degeneracy at the $\Phi_A-\Phi_S$ boundary is interpreted, from this vantage point, by the fact
that hosts 2 do not accommodate hybrids 5 alone.

By analogous reasoning based on the transcription (\ref{eq:2-8})  we can cast Eqs.~(\ref{eq:3-2}) for the 
particles excited from the plateau state (Sec.~\ref{sec:ppv}) in the form (\ref{eq:3-5}) with the right-hand 
sides changed into
\begin{align}\label{eq:3-10} 
 & e^{-2K_J},\quad e^{-K_J-K_D-H},\quad e^{-K_J-K_D+H}, \nonumber \\
 & e^{4K_J},\quad e^{-3K_J},\quad 1,
\end{align}
and the associated grand partition function in the form
\begin{equation}\label{eq:3-11} 
Z=\Big[(1+w_3)e^{-K_J-K_D/2-3H/2}\Big]^N.
\end{equation}

The calculation of the $\langle \bar{N}_m\rangle$ produces the expected results.
The region with a low density of particles of any kind at low $T$ is now $\Phi_{P+}$.
Regions $\Phi_{F+}, \Phi_S$ have high densities of compacts $m=1,2$, respectively,
region $\Phi_{F-}$ has a high density of tags $m=3$, and regions $\Phi_A, \Phi_{P-}$ have
high densities of hosts $m=4,5$, respectively.
In this case the low-degeneracy of the $\Phi_A-\Phi_S$ phase boundary is explained by the 
observation that the negative-energy compacts $m=2$ and the negative-energy hosts $m=4$ 
do not interlink directly, only via (zero-energy) elements of pseudo-vacuum.

From the expressions for $Z$ and the $\langle \bar{N}_m\rangle$ (all in four versions) a great many 
statistical mechanical results can be derived.
However, the focus of this paper is on the the identification of functions associated with 
particles of given structures in specific environments.

%
\section{Entropy of mixing}\label{sec:entmix}  
%
Depending on whether or not the pseudo-vacuum of choice is a state of lowest energy, the Ising chain
(\ref{eq:1-1}) inside any of the six regions in the parameter space  of Fig.~\ref{fig:phadah}(a)
can be interpreted (at low $T$) either as a dilute gas of several species of (positive-energy) particles
or as a liquid of predominantly one species of (negative-energy) particles with low concentrations of 
higher-energy particles added.
The physical processes that take place when the Hamiltonian parameters are gradually changed across 
one or the other phase boundary allow for multiple interpretations in terms of particles from the four sets
$\{m_A\}$, $\{m_S\}$, $\{m_{F+}\}$, $\{m_{P+}\}$, 
introduced in Sec.~\ref{sec:combi} and elements of the associated pseudo-vacua.

We now consider the simplified phase diagram pertaining to $h=0$ as shown in Fig.~\ref{fig:phadah}(b).
The singlet phase $\Phi_S$ and the antiferromagnetic phase $\Phi_A$ are the pseudo-vacua of particles 
from the sets $\{m_S\}$ and $\{m_A\}$, respectively.
The ferromagnetic phase $\Phi_F$ represents the degenerate  phases $\Phi_{F\pm}$.
One ground-state vector is the pseudo-vacuum for particles from set $\{m_{F+}\}$ and the other is a solid
of zero-energy particles from that set.
The plateau phases $\Phi_{P\pm}$ are not realized at $h=0$.

Along the two phase boundaries at $D>0$, the degeneracy of the ground state is low (threefold)  and 
along the third phase boundary (at $D<0$) it is exponentially high.
This qualitative difference has a natural explanation expressible in terms of particle functions.

\subsection{Surfactants}\label{sec:surf}  
We begin by examining the physics near the $\Phi_A-\Phi_S$ boundary simultaneously from two vantage 
points.
Phase $\Phi_A$ is the pseudo-vacuum of particles from set $\{m_A\}$ or a solid of hybrids $m_S=7$
$(\uparrow\downarrow\uparrow,\downarrow\uparrow\downarrow)$.
Conversely, phase $\Phi_S$ is the pseudo-vacuum of particles from set $\{m_S\}$ or a solid of tags 
$m_A=3$ $(\circ\circ)$.
At the phase boundary, the lowest excitations are separated by $\Delta E=J$ from the threefold ground 
state and contain exactly one host particle acting as a pair of barriers between one segment of tags or 
hybrids on the inside and one segment of pseudo-vacuum on the outside.
Viewed from one vantage point, the host is among $m_A=4,5,6$ 
$(\uparrow\circ\uparrow, \downarrow\circ\downarrow, \uparrow\circ\downarrow, 
\downarrow\circ\uparrow)$ and accommodates tags $m_A=3$.
The other vantage point has one host $m_S=1,2,3,4$  $(\circ\uparrow\circ, \circ\downarrow\circ, 
\circ\uparrow\downarrow\circ, \circ\downarrow\uparrow\circ)$ accommodating hybrids $m_S=7$.

When we switch vantage points the host particle is turned inside out and the tags or hybrids inside
become elements of pseudo-vacuum outside. 
One product state of this kind is
\begin{align}\label{eq:4-1} 
& |~\underbrace{\cdots\uparrow\downarrow}_{\mathrm{vac.}}
\underbrace{\uparrow\circ\circ\cdots\circ\circ\downarrow}_{\mathrm{host+tags}}
\underbrace{\uparrow\downarrow\cdots}_{\mathrm{vac.}}~\rangle \nonumber \\
&= |~\underbrace{\cdots\circ\circ}_{\mathrm{vac.}}
\underbrace{\circ\downarrow\uparrow\downarrow\cdots\uparrow\downarrow\uparrow
\circ}_{\mathrm{host+hybrids}}
\underbrace{\circ\circ\cdots}_{\mathrm{vac.}}~\rangle
\end{align}
with host $m_A=6$ and tags $m_A=3$ embedded in the antiferromagnetic pseudo-vacuum 
or host $m_S=4$ and hybrids $m_S=7$ embedded in the singlet pseudo-vacuum.

From either vantage point the host assumes the function of a surfactant between elements of vacuum on 
one side and tags or hybrids on the other side.
The absence of direct links between tags/hybrids and elements of vacuum implies an infinite surface 
tension.
The presence of surfactants reduces it to a finite value.

At the phase boundary, the energy densities inside and outside the host are equal.
Moving away from the phase boundary raises the energy density of one segment and lowers that of the 
other, making the growth of one segment at the expense of the other energetically favorable.
This process involves either the creation of negative-energy particles or the annihilation  of
positive-energy particles from the tag or hybrid categories.

To get the transition started, one host particle (of positive energy) must first be created.
Its energy at the phase boundary is equal to $J$ and decreases on a path into the phase with opposite
pseudo-vacuum. Eventually it goes negative, producing a hysteresis effect.

What happens when the phase boundary $D=J$ is crossed at low but nonzero $T$ evokes a 
metamorphosis reminiscent of an M.~C.~Escher print. 
As the tags $m_A=3$ soften energetically, their population explodes.
The hosts of the soft tags expand in the process.
The number of thermally excited hosts remains low.
The segments of pseudo-vacuum between the hosts shrink to sizes comparable to the expanded hosts.

A change of perception identifies the close-packed soft tags of set $\{m_A\}$ as segments of
pseudo-vacuum for set $\{m_S\}$ and segments of pseudo-vacuum for set $\{m_A\}$ as close-packed
soft hybrids of set $\{m_S\}$.
The hosts from set $\{m_A\}$ break apart and reassemble into hosts from set $\{m_S\}$.
Each left half of an old host pairs up with the right half of the closest old host to its left to form a new 
host.

\subsection{Immiscible liquids}\label{sec:immisc}  
Now we view the physics near the same $\Phi_A-\Phi_S$  boundary from a third vantage point.
We describe it in terms of particles from the set $\{m_{F+}\}$ whose pseudo-vacuum is not a 
lowest-energy state.
Each of the three vectors that make up the physical vacuum at the phase boundary is a solid of one 
species of negative-energy particles: two vectors of hosts $m_{F+}=2$ 
$(\uparrow\downarrow\uparrow)$ and one vector of hybrids $m_{F+}=5$ $(\circ\circ)$.
The energy per bond is $-2J$ for both species.
Hosts $m_{F+}=2$ do not accommodate hybrids $m_{F+}=5$.

The lowest excited states contain one host $m_{F+}=3,4$ 
$(\uparrow\circ\downarrow\uparrow, \uparrow\downarrow\circ\uparrow)$ 
with energy per bond $-5J/3$ at the phase boundary.
These hosts interlink externally with hosts $m_{F+}=2$ and internally with hybrids $m_{F+}=5$. 
They assume the function of surfactant by reducing the interfacial tension from infinity to a finite value.
Varying the parameters $J,D$ across the phase boundary tips the energy balance from favoring hosts
$m_{F+}=2$ to hybrids $m_{F+}=5$ or vice versa.
Any transition process that involves local spin flips cannot take place without  the presence of
a host $m_{F+}=3,4$, which introduces the hysteresis effects noted previously.

At low, nonzero $T$ the thermodynamic state near the $\Phi_A-\Phi_S$ phase boundary exhibits 
attributes of two immiscible liquids.
There are two species of particles with high population densities, the hosts $m_{F+}=2$ and the hybrids 
$m_{F+}=5$.
They coexist in an emulsion of sorts with higher-energy hosts $m_{F+}=3,4$ of much lower density 
acting as emulsifiers.

\begin{figure}[t]
  \includegraphics[width=85mm]{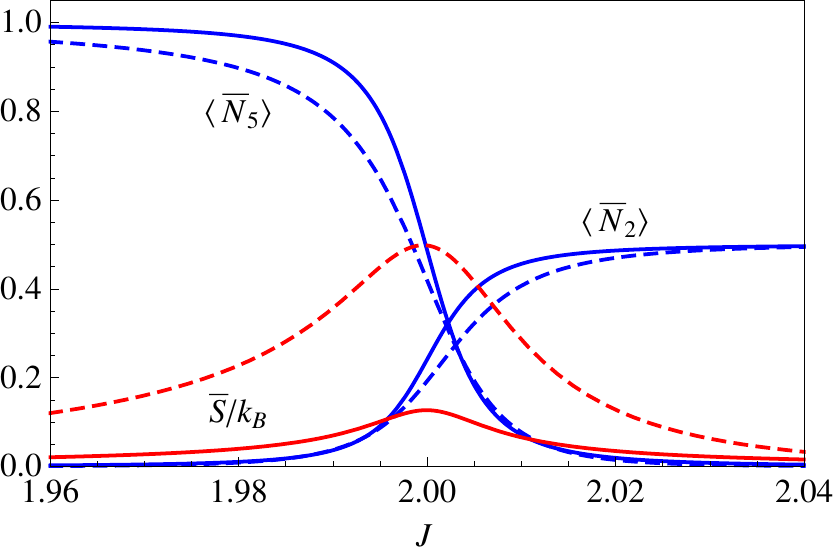}
\caption{Average numbers (per site) $\langle \bar{N}_{2}\rangle$, 
  $\langle\bar{N}_{5}\rangle$ from the set $\{m_{F+}\}$ and entropy (per site) $\bar{S}/k_B$
  versus $J$ with $D=4-J$ at $k_BT=0.25$ (solid lines) and  $k_BT=0.5$ (dashed lines) in arbitrary 
  energy units. The scale on the horizontal axis is for the lower $T$. The curves at the higher $T$ are for 
  $1.0<J<3.0$. }
  \label{fig:dlj2-phif-as12}
\end{figure}

In Fig.~\ref{fig:dlj2-phif-as12} we have plotted (at two fixed values of $T$) the population densities
$\langle\bar{N}_2\rangle$, $\langle\bar{N}_5\rangle$ of the particles that represent the two liquids
versus $J$ along a line that crosses the $\Phi_A-\Phi_S$ boundary in a perpendicular direction.
Also shown is the entropy per site as derived from (\ref{eq:3-9}). 

Away from the phase boundary we have a fairly pure liquid of either hosts $m_{F+}=2$ or tags 
$m_{F+}=5$.
The different number densities are caused by the different particle sizes.
Particles of both liquids coexist in comparable numbers near the phase boundary $(J=2)$.

The entropy has a maximum at the phase boundary.
The key observation is that as $T$ is lowered the coexisting populations of particles representing the 
two liquids remain high but the entropy is much reduced.
It can be interpreted as the entropy of mixing of two immiscible liquids \cite{note6}.

The physics near the $\Phi_S-\Phi_F$ boundary described from the vantage points of the four sets 
of particles is a variation of the same theme.
One significant difference is that the thermodynamic state as viewed from the vantage point of the 
particle set $\{m_A\}$ involves three immiscible liquids rather than just two.

\begin{figure}[t]
  \includegraphics[width=85mm]{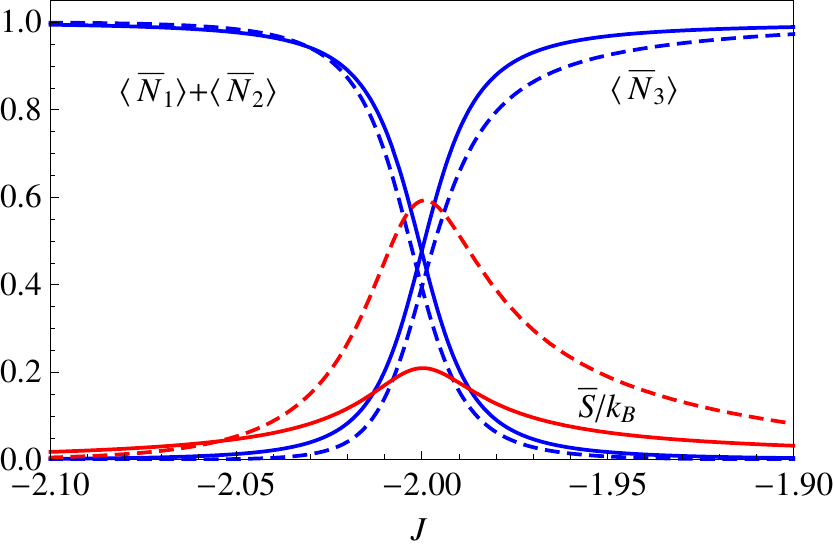}
\caption{Average numbers (per site) $\langle \bar{N}_{1}\rangle+\langle \bar{N}_{2}\rangle$, 
  $\langle\bar{N}_{3}\rangle$ from the set $\{m_A\}$ and entropy (per site) $\bar{S}/k_B$
  versus $J$ with $D=4+J$ at $k_BT=0.3$ (solid lines) and  $k_BT=0.6$ (dashed lines) in arbitrary 
  energy units. The scale on the horizontal axis is for the lower $T$. The curves at the higher $T$ are for 
  $-3.0<J<-1.0$.}
  \label{fig:dlj2-phia6-sf12}
\end{figure}

The particle species with high population densities at low $T$ are the compacts $m_A=1,2$ 
$(\uparrow\uparrow, \downarrow\downarrow)$ at $J<-D$ and the tags $m_A=3$ $(\circ\circ)$ at $J>-D$ 
as shown in Fig.~\ref{fig:dlj2-phia6-sf12}. 
Throughout region $\Phi_F$ the negative-energy compacts remain segregated by higher-energy 
hosts $m_A=6$ $(\uparrow\circ\downarrow, \downarrow\circ\uparrow)$ or elements of 
pseudo-vacuum $(\uparrow\downarrow, \downarrow\uparrow)$.
The entropy is very low.
At the phase boundary the compacts coexist with tags $m_A=3$, but all three species remain 
segregated.
The tags exist inside hosts and the compacts outside.

In addition to hosts $m_A=6$, hosts $m_A=4,5$ $({\uparrow\circ\uparrow,} 
{\downarrow\circ\downarrow})$ 
are present in comparable numbers.
The particle functions that segregate compacts from each other are different from those that
segregate compacts from tags.
The shallow peak in the entropy curve at low $T$ is a dim signal of the increase in emulsion 
components near the phase boundary.

\subsection{Miscible liquids}\label{sec:misc}  
The physics near the $\Phi_A-\Phi_F$ phase boundary is qualitatively different.
From the vantage points of the adjacent phases the two vectors of phase $\Phi_A$ 
are the pseudo-vacuum of $\{m_A\}$, composed of elements 
$\uparrow\downarrow,\downarrow\uparrow$, or two solids composed of hosts $m_{F+}=2$ 
$(\uparrow\downarrow\uparrow)$, whereas the two vectors of phase $\Phi_F$ are the pseudo-vacuum 
of $\{m_{F+}\}$, composed of elements $\uparrow\uparrow$,
plus a solid of tags $m_{F+}=6$ $(\downarrow\downarrow)$ or two solids, one composed of compacts $m_A=1$ $(\uparrow\uparrow)$ and the other of compacts $m_A=2$ $(\downarrow\downarrow)$.

The low degeneracy of the pure phases is explained by the fact that the particles and elements of 
pseudo-vacuum involved in each pair of vectors do not mix.
However, when both pairs of vectors become degenerate (at the phase boundary) the options for 
particles or elements of pseudo-vacuum to interlink mutually increase dramatically, producing a 
$2^N$-fold degeneracy, which comprises all vectors $|\sigma_1\cdots\sigma_N\rangle$ with 
$\sigma_l=\uparrow,\downarrow$.
Any such state can be interpreted as either a combination of compacts $m_A=1,2$ 
$(\uparrow\uparrow,\downarrow\downarrow)$ and elements of associated pseudo-vacuum 
$(\uparrow\downarrow,\downarrow\uparrow)$ or as a combination of hosts $m_{F+}=2$
$(\uparrow\downarrow\uparrow)$, tags $m_{F+}=6$ $(\downarrow\downarrow)$, and 
elements of associated pseudo-vacuum $(\uparrow\uparrow)$.

From the vantage point of particle set $\{m_S\}$, the physical vacuum at the $\Phi_A-\Phi_F$ phase
boundary consists close-packed configurations of particles with the lowest negative energy per bond.
They are the hybrids $m_S=5,6,7$ 
$(\uparrow\uparrow;\downarrow\downarrow;\uparrow\downarrow\uparrow,
\downarrow\uparrow\downarrow)$.
Their ability to interlink produces all $2^N$ vectors of the ground state at the phase boundary.
Varying the parameter $J$ across the phase boundary at fixed $D<0$ enables the transition to proceed 
via local spin-flip processes without energy barrier.
There are no hysteresis effects in this case.

At low, nonzero $T$ the thermodynamic state near the phase boundary can be interpreted as that of 
three liquids.
The species of particles with high population densities are the hybrids $m_S=5,6$ at $J<0$ and the 
hybrids $m_S=7$ at $J>0$.
The state in phase region $\Phi_F$ is akin to that of two immiscible liquids as already
described (in Sec.~\ref{sec:immisc}) from the vantage point of set $\{m_A\}$. 
However, here the two relevant species are hybrids.
In region $\Phi_A$, by contrast, we have a one-component liquid, composed of hybrids $m_{S}=7$.
The crucial difference in this scenario compared to the one described near the $\Phi_S-\Phi_F$ phase
boundary is that the hybrids $m_{S}=7$ not only interlink with both species of hybrids, $m_{S}=5,6$,
but they also mediate their mixing by acting as a surfactant with zero interfacial energy.

\begin{figure}[h]
  \includegraphics[width=85mm]{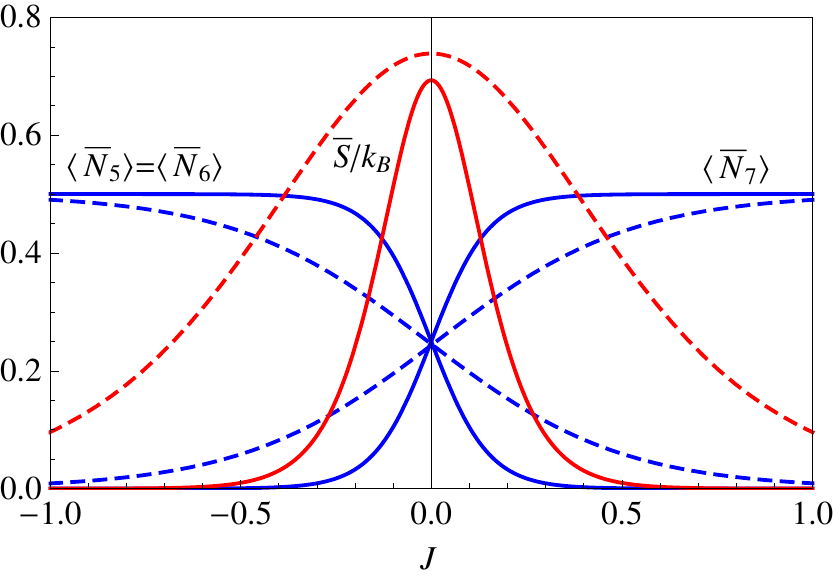}
\caption{Average numbers (per site) $\langle \bar{N}_{5}\rangle=\langle \bar{N}_{6}\rangle$, 
  $\langle\bar{N}_{7}\rangle$ from the set $\{m_S\}$ and entropy (per site) $\bar{S}/k_B$
  versus $J$ at $D=-2$  at $k_BT=0.15$ (solid lines) and  $k_BT=0.5$ (dashed lines) in arbitrary 
  energy units. The scale on the horizontal axis is the same for all curves.}
  \label{fig:dlj2-phis-af12}
\end{figure}

Figure~\ref{fig:dlj2-phis-af12} shows the population densities of hybrids $m_S=5,6,7$ near the 
$\Phi_A-\Phi_F$ phase boundary and also the entropy, all at two different temperatures.
We observe a strikingly different behavior of the entropy curve in comparison to its behavior across the 
other two phase boundaries (shown in Figs.~\ref{fig:dlj2-phif-as12}, \ref{fig:dlj2-phia6-sf12}). 
Here the particles with dominant populations behave as in miscible liquids. 
Again there is a region where these populations coexist in comparable high 
densities at both the higher and the lower temperature. 
Unlike in the previous two scenarios the entropy stays high at the lower temperature because the particle 
populations are not separated by barriers in the form of higher-energy host particles acting as surfactant 
with significant interfacial energies.

%
\section{Structure and function}\label{sec:strufunc}  
%
In this project we started out with interacting Ising spins on a periodic chain.
The model Hamiltonian (\ref{eq:1-1}) causes the self-assembly of particles with various structures
that (i) remain free of interaction energies even at high densities
and (ii) are robust against changes in the Hamiltonian parameters.

We have identified four sets of particle species excited from four different ordered Ising states.
From there on we are free to remove the scaffold of the Ising model and assign  arbitrary
energies to all particles species \cite{note10}.

Each particle species from a given set has a distinct structure.
That structure is encoded in one or several motifs.
The structure of a particle is only meaningful in relation to the structure of the associated 
pseudo-vacuum.
Particles $m_A=2$, $m_S=6$, and $m_{P+}=3$, for example, all have the same motif, 
$\downarrow\downarrow$, but represent different structures. 
They belong to different species, excited from different pseudo-vacua.
The statistical interaction specifications of all three species are different.

We have classified all particle species into four categories according to structural features.
These features represent specific particle functions.
Compacts have only external links.
They interlink with elements of pseudo-vacuum and with external links of compacts and hosts.
Hosts have external and internal links.
They interlink externally with elements of pseudo-vacuum and with compacts and hosts.
They interlink internally with tags or hybrids.
Tags and hybrids interlink with internal links of hosts and, selectively, with each other.
These limited options for particles to interlink mutually may be interpreted as a manifestation
of stereospecific binding with zero binding energy.

The capacities of hosting and being hosted are functions that some particle species have 
and other species lack.
Some functions of particle species depend on the circumstances as is the case in two
examples encountered in Sec.~\ref{sec:entmix}.
(i) In a situation where two particle species that do not interlink directly have equal negative energy 
density, lower than the energy density of any other species, then they form two immiscible liquids a low 
$T$.
Segments of one liquid are separated from segments of the other.
The function of surfactant is assumed by a third particle species or by elements of pseudo-vacuum
with higher density.
(ii) If the circumstances change in such a way that the energy density of the surfactant particle species 
becomes equal to that of the particle species representing the two immiscible liquids then the surfactant 
particle changes its function into a miscibility agent of sorts.

Modifications in the model Hamiltonian (\ref{eq:1-1}) affect the nature of particles.
We have already seen that variations in the Hamiltonian parameters $J,D,h$ cause changes in 
some functions of particles even though their structures remain invariant.

Additional commuting terms in the Ising Hamiltonian such as a next-nearest-neighbor coupling
or a bond alternation make it necessary to find new sets of particles with longer motifs that interlink 
differently and exhibit different statistical interactions.
Ising Hamiltonians with spin $s>1$ pose similar challenges apart from the need for motifs
with new symbols \cite{note9}.

Non-commuting terms in the Hamiltonian are apt to affect the particles introduced here 
more drastically and in multiple ways.
The particles are likely to move along the lattice, scatter off each other, and decay into other particles.
Under special circumstances as mentioned in Sec.~\ref{sec:intro}, the particles do not decay and 
experience only elastic two-body scattering.
Such particles can be characterized by motifs in momentum space \cite{LMK09}.
A taxonomy of particles according to structures and functions in this wider arena is, of course, a more 
challenging project.

%
%



\begin{thebibliography}{100}
\bibitem{note1} 
For $J<0$ only the phases $\Phi_S, \Phi_{F\pm}$ exist.
The twofold ground states associated with phases $\Phi_A, \Phi_{P\pm}$ are only realized for even $N$.
\bibitem{note7}
Hamiltonian terms that assign different energies to the two motifs are then permissible.
\bibitem{note3}
Some evidence for $s=\frac{1}{2}$ is implicit in the results reported in Ref.~\cite{LVP+08}.
Further evidence will be presented in forthcoming studies that focus on different aspects
of the statistical mechanics of particles with shapes.
\bibitem{note2}
The analysis assumes that $N\gg1$ and $\langle N_m\rangle\gg1$.
Any part of a nonzero $A_m$ that is of O(1) is ignorable.
Our choice of reference energy for the $\epsilon_m$ implies zero chemical potential
for all particle species.
\bibitem{note4}
The transfer matrix solution of (\ref{eq:1-1}) leads to the same cubic equation.
The canonical partition function inferred from the largest eigenvalue of the transfer matrix
is identical to $Z$ if the same energy scales are used.
\bibitem{note8}
It is conceivable that the physical vacuum in some phase region is formed by a regular 
pattern formed by more than one species but that possibility is not realized here.
\bibitem{note5}
Partition functions (\ref{eq:3-6}) and (\ref{eq:43-3}) differ by a factor $e^{\Delta E/k_BT}$, 
where $\Delta E$ is the energy difference between the two pseudo-vacua used.
This makes it possible to express the $w_m$ in terms of the dominant eigenvalue of 
the transfer matrix.
\bibitem{note6}
The same physics at the $\Phi_A-\Phi_S$ is described if we switch the vantage point to the plateau 
state and the particle set $\{m_{P+}\}$. Only the protagonists change.
\end{thebibliography}
\end{document}